\documentclass[12pt,onecolumn,draftcls]{IEEEtran}

\usepackage{amsmath}
\usepackage{stmaryrd}
\usepackage{graphicx}
\usepackage{fullpage}

\title{Outage Probability of Multiple-Input Single-Output (MISO) Systems with
Delayed Feedback$^{\dagger}$\footnote{$^{\dagger}$This work was
done at the Dept. of Electrical Engg., Indian
Institute of Technology Madras. Part of this work was presented
at the $40^{th}$ Asilomar Conference on Signals, Systems, and
Computers held in Oct-Nov 2006 at Pacific Grove, CA.}}

\author{Venkata Sreekanth Annapureddy$^{1}$, Devdutt V. Marathe$^{2}$, T. R. Ramya$^{3}$ and Srikrishna 
Bhashyam$^{3}$\\ ~\\ \vspace*{-5mm}
$^{1}$ Coordinated Science Laboratory\\ Department of Electrical and Computer Engineering\\ University of Illinois at Urbana-Champaign\\ 1308 West Main St. Urbana, IL 61801\\  E-mail: {\em vannapu2@uiuc.edu} \\ ~ \\ \vspace*{-5mm}
$^{2}$ Indian Institute of Management\\ Vastrapur, Ahmedabad 380015.\\ 
E-mail: {\em devdutt.marathe@gmail.com} \\ ~ \\ \vspace*{-5mm}
$^{3}$Department of Electrical Engineering\\
Indian Institute of Technology Madras \\ Chennai 600036, India\\
Phone: 91-44-22574439, Fax: 91-44-22570120\\ E-mail: {\em
\{ee04d016,skrishna\}@ee.iitm.ac.in} \\
}

\begin{document}
\maketitle

\newpage
\begin{abstract}
We investigate the effect of feedback delay on the outage probability
of multiple-input single-output (MISO) fading channels. Channel state
information at the transmitter (CSIT) is a delayed version of the
channel state information available at the receiver (CSIR). We
consider two cases of CSIR: (a) perfect CSIR and (b) CSI estimated at
the receiver using training symbols. With perfect CSIR, under a
short-term power constraint, we determine: (a) the outage probability
for beamforming with imperfect CSIT (BF-IC) analytically, and (b) the
optimal spatial power allocation (OSPA) scheme that minimizes outage
numerically. Results show that, for delayed CSIT, BF-IC is close to
optimal for low SNR and uniform spatial power allocation (USPA) is
close to optimal at high SNR. Similarly, under a long-term power
constraint, we show that BF-IC is close to optimal for low SNR and
USPA is close to optimal at high SNR. With imperfect CSIR, we obtain
an upper bound on the outage probability with USPA and BF-IC. Results
show that the loss in performance due to imperfection in CSIR is not
significant, if the training power is chosen appropriately.
\end{abstract}

\vspace{-6mm}
\section{Introduction}

Channel State Information is very crucial in determining the
performance of any wireless system. The minimum outage probability of
multiple-input single-output (MISO) channels with perfect channel
state information at the receiver (CSIR) and no channel state
information at the transmitter (CSIT) is derived in
\cite{telatar}. For reasonably low outage probabilities, uniform
spatial power allocation (USPA) across the spatial dimension is the
optimal strategy. Outage probability of MISO systems with perfect CSIT and
CSIR is derived in \cite{BigCaiTar}. It is shown that feeding back the
CSI provides significant gain in the performance, and that beamforming
to the direction of the channel is optimal and provides a constant SNR
gain over no CSIT under short-term power constraint (i.e., transmit
power is constant over each transmission interval). In the case of
long-term average power constraint, it is also possible to adapt the
transmission power level based on channel feedback (i.e., temporal
power control). Outage can be reduced significantly by saving power
when the channel is strong and using the saved power when the channel
is worse. The optimum power allocation strategy to minimize the outage
probability over fading channels and MISO fading channels is
determined in \cite{gcp} and \cite{BigCaiTar} respectively.

In practice, the feedback channel resources are seldom perfect enough
to provide instantaneous and noiseless feedback. Under the short term
power constraint, and for two cases of imperfect feedback namely mean
feedback and covariance feedback, spatial schemes that a) minimize the
outage probability are studied in \cite{costas,arie}, and b) maximize
the mutual information are studied in \cite{madhow}. In \cite{dabak},
BER performance of spatial schemes in the presence of delayed feedback
has been studied. Under a long-term power constraint, minimum outage
probability with temporal power control for quantized CSIT has been
studied in \cite{sk-ba}. In practice, it is also not feasible to have
a perfect estimate of CSIR. Usually, channel state information at the
receiver is estimated using training symbols, and the resources used
during the training period have to be accounted for. Outage
probability with preamble based CSIR and quantized CSIT has been
studied in \cite{sk-ba}. In \cite{yoo-ag}, maximizing mutual
information in the presence of channel estimation error and delayed
feedback has been studied.

In this paper, we focus on the effect of the delay in feedback on the
performance from the point of view of outage probability. Using the
delayed feedback model in \cite{macro}, we solve the problem of
minimum outage transmission over MISO channels under both short-term
and long-term power constraints. Under a short-term power constraint,
beamforming is optimal if the transmitter has perfect CSI. We analyze
the loss in performance of beamforming due to the delay in the
feedback and derive an analytical expression for the outage
probability of beamforming with imperfect CSIT (BF-IC). Results show
that BF-IC, which allocates total power in the direction of CSIT, is
better at low SNR while USPA \cite{telatar}, which allocates equal
power in all the directions and does not require any feedback, is
better at high SNR. However, none of the above two strategies is
optimal. The minimum outage transmission strategy for a given delay,
optimal spatial power allocation (OSPA) is determined. OSPA involves
beamforming along the spatial modes and optimal power allocation
across the spatial modes. Numerical results show that BF-IC is very
close to OSPA for low SNR while USPA is close to OSPA for high
SNR. Since OSPA does not provide significant gain at any SNR, compared
to the best of BF-IC and USPA, the cross-over SNR at which USPA
becomes better than BF-IC is important and can be used to switch
between BF-IC and USPA. We present the equation to determine this
cross-over SNR and solve it numerically.

Under a long-term power constraint, with perfect CSIT, the optimal
beamforming (to the channel direction) and temporal power control
strategy is obtained in \cite{BigCaiTar}. We numerically evaluate the
outage probabilities for BF-IC and USPA with temporal power
control. Again, BF-IC is better at low SNR while USPA is better at
high SNR. Finally, we extend the analysis with delayed feedback and
perfect CSIR to the case of delayed feedback and imperfect CSIR. An
upper bound on the outage probability of USPA and BF-IC with imperfect
CSIR is obtained. The loss in performance due to the error in
estimation of CSIR is shown to be negligible if the training power is
chosen optimally.

The rest of the paper is organized as follows. In Section \ref{model},
our system model is introduced. In Section \ref{stpc}, under the short
power constraint, outage probability with BF-IC and OSPA are determined
and compared with USPA. In Sections \ref{ltpc} and \ref{imcsir}, the
long term power constraint and imperfect CSIR are considered. Finally,
Sections \ref{results} and \ref{conc} present the results and conclusions.

\vspace*{-4mm}
\section{System Model}
\label{model}
The MISO system with $M$ transmit antennas and 1 receive antenna is,
as usual, modeled as
\begin{equation}
y = \textbf{h}^{H}\textbf{x} + z,
\end{equation}
\noindent where $\textbf{h} \sim \cal{CN} (\bf{0},\bf{I})$ is a
$M\times 1$ independent, identically distributed (i.i.d) and zero-mean
circularly symmetric complex Gaussian channel vector, $\textbf{x}$ is
a $M\times 1$ channel input vector and $z$ is zero-mean unit-variance
additive white Gaussian noise (AWGN). We use a block fading model, where
the channel coefficients are assumed to be fixed within a given duration,
known as coherence interval. We assume high correlation between successive
time durations. Using the Gaussian channel vector model, the delay in the 
feedback is captured by the correlation coefficient $\rho{}$ between CSIT 
and CSIR. The old channel and the actual channel can be  related as
follows \cite{macro}:
\begin{equation}
\textbf{h} = \rho \textbf{h}_{old} + \sqrt{1-\rho ^{2}}\textbf{w},
\label{h-hold}
\end{equation}
\noindent where $\textbf{h}_{old}$ is the delayed CSIT, $\rho$ is a
correlation coefficient, and $\textbf{w} \sim  \cal{CN} (\bf{0},\bf{I})$ is independent of $\textbf{h}_{old}$. The gap between no CSIT ($\rho = 0$) and perfect CSIT ($\rho = 1$) is bridged using $\rho$. Lower the delay in the feedback, higher the value of $\rho$. 

\section{Short-Term Power Constraint}
\label{stpc}
Assuming a short-term power constraint \cite{gcp}, such that the transmit
power is not a function of time, the mutual information is given by
\begin{equation}
 I(\textbf{x};y/\textbf{h,h}_{old}) = \textrm{log} ( 1 +
 P\textbf{h}^H\textbf{Q}\textbf{h} ),
\label{I-Q}
\end{equation}
\noindent where $\textbf{Q}$ is the input covariance matrix such that 
$Tr(\textbf{Q}) = 1$ and $P$ is the transmit power. 

Consider the two extreme cases: zero feedback ($\rho = 0$) and instantaneous feedback
($\rho = 1$). For $\rho=0$, where the transmitter does not have any 
knowledge of the channel state information, the diversity strategy
with the power distributed equally among the $M$ orthogonal
independent transmit directions, i.e., USPA is optimal \cite{telatar}, i.e., we have $\displaystyle{\textbf{Q} = \frac{\textbf{I}}{M}}$, and
\begin{equation}
\begin{split}
{P_{out}}_{\textbf{USPA}}(M,R,P) & = \Gamma{}_M\left( \frac{e^R -
1}{P/M}\right),
\label{pout_uspa}
\end{split}
\end{equation}
where ${P_{out}}_{\textbf{USPA}}(M,R,P)$ is the outage probability (as
defined in \cite{telatar}) for a $M \times 1$ system using USPA
corresponding to a transmit power constraint $P$ and rate $R$ (in
nats/transmission), and $\Gamma_M (\cdot)$ is the incomplete Gamma
function defined as 
$\displaystyle{\Gamma_M(x) = \frac{1}{(M-1)!}\int_0^{x} t^{M-1} e^{-t} dt}$.

For $\rho = 1$, where the transmitter has perfect CSI, beamforming is
optimal \cite{BigCaiTar}, i.e., $\displaystyle{{\bf x} = \frac{{\bf
h}}{\sqrt{{\bf h}^H{\bf h}}}s}$, where $s$ is a scalar i.i.d. Gaussian
input, $\displaystyle{\textbf{Q} =
\frac{\textbf{h}\textbf{h}^H}{\textbf{h}^H\textbf{h}}}$, and
\begin{equation}
\begin{split}
{P_{out}}_{\textbf{BF}}(M,R,P,\rho = 1) = \Gamma{}_M\left( \frac{e^R -
1}{P}\right).
\end{split}
\label{bf1}
\end{equation}
Outage performance for $\rho = 1$ is 10log$_{10}$M dB better than the
performance for $\rho = 0$. For $0 < \rho{} < 1$, where we do not have
perfect CSIT, we evaluate the outage performance of beamforming using
the imperfect CSIT in Section \ref{bficperfect}. We also determine the
optimal spatial power allocation strategy that minimizes the outage
probability in Section \ref{ospa-sec} and compare it with beamforming
using the imperfect CSIT.

\vspace*{-6mm}
\subsection{Beamforming using imperfect CSIT (BF-IC)} 
\label{bficperfect}
In this section, the loss in performance due to the presence of the
delay in the feedback is analyzed and an expression for the outage
probability (equation (\ref{result_bf})) is derived. This is a simple
extension of beamforming from perfect CSIT to the imperfect CSIT case,
where beamforming is performed using the imperfect CSIT assuming that
it is the actual channel. Therefore, we have $\displaystyle{{\bf x} =
\frac{{\bf h}_{old}}{\sqrt{{\bf h}_{old}^H{\bf h}_{old}}}s}$, where
$s$ is a scalar i.i.d. Gaussian input, and
\begin{equation}
\textbf{Q} = \frac{\textbf{h}_{old}\textbf{h}^H_{old}}{\textbf{h}^H_{old}\textbf{h}_{old}}.
\label{Q-bf}
\end{equation}
Substituting (\ref{Q-bf}) in (\ref{I-Q}) and denoting the feedback SNR $\textbf{h}^H_{old}\textbf{h}_{old}$ by $\gamma{}$, we get
\begin{equation}
I(\textbf{x};y/\textbf{h,h}_{old}) = \textrm{log} \left(1 +
P\frac{\textbf{h}^H\textbf{h}_{old}\textbf{h}^H_{old}\textbf{h}}{\gamma{}}\right).
\end{equation}
Now, we derive the outage probability for the specific model described
in equation (\ref{h-hold}).  Note that $\gamma{}$ is Gamma distributed
with the pdf given by
\begin{equation}
f_{\Gamma{}}(\gamma{}) = \frac{\gamma{}^{M-1}e^{-\gamma{}}}{(M-1)!}.
\end{equation}
The expression for the mutual information for a given
$\textbf{h}_{old}$ can be simplified as follows. 
\begin{equation}
\begin{split}
 \frac{\textbf{h}^H\textbf{h}_{old}\textbf{h}_{old}^H\textbf{h}}{\gamma} =  \frac{ | \textbf{h}^H\textbf{h}_{old}|^{2}}{\gamma} = \frac{ | (\rho \textbf{h}_{old} + \sqrt{1-\rho ^{2}}\textbf{w} )^H\textbf{h}_{old}|^{2}}{\gamma}\\
 = \frac{(1-\rho ^{2}) }{2}\left| \sqrt{\frac{2 \rho^{2}}{(1-\rho ^{2})}\gamma}  +  \sqrt{2}\frac{\textbf{w}^H\textbf{h}_{old}}{\sqrt{\gamma}} \right|^{2}.
 \end{split}
\end{equation} 
Hence, the mutual information given $\textbf{h}_{old}$ can be simplified as
\begin{equation}
I(\textbf{x};y/\textbf{h,h}_{old}) =  \textrm{log} \left( 1 + P \frac{(1-\rho ^{2}) }{2}A \right), 
\end{equation}
\noindent where $A = \left| \sqrt{\delta} +
\sqrt{2}\frac{\textbf{w}^H\textbf{h}_{old}}{\sqrt{\gamma}} \right|^2$,
$\delta = 2\mu\gamma$ and $\mu = \frac{\rho^2}{1-\rho^2}$. Note that $\frac{\textbf{w}^H\textbf{h}_{old}}{\sqrt{\gamma}}$ given $\textbf{h}_{old}$ is a zero mean complex Gaussian random variable with variance $ || \frac{\textbf{h}_{old}^H}{\sqrt{\gamma}} ||^2  = 1$. Thus, $A$ given $\gamma$ is a non-central chi-square $(\text{nc-}\chi^2)$
random variable with two degrees of freedom and parameter $\delta
$. Observe that the distribution of mutual information given $\textbf{h}_{old}$ depends on only $\gamma = |\textbf{h}_{old}|^2$. Therefore we have the following expression for the outage
probability for a given $\gamma$.
\begin{equation}
\begin{split}
\mbox{Pr}(\mbox{outage}/\gamma) =& \mbox{Pr}\left(\textrm{log} \left( 1 + P \frac{(1-\rho ^{2}) }{2}A \right) < R\right)\\
=& F_{(\text{nc-}\chi^2,2,\delta)}(2\beta{}),
\end{split}
\label{pout-bf-gamma}
\end{equation}
\noindent where $\beta=\frac{e^R - 1}P(\mu+1)$, and 
$F_{(\text{nc-}\chi^2,2,\delta)}(\cdot)$ is the CDF
of a non-central chi-square random variable with two degrees of
freedom and parameter $\delta$. The overall probability of outage can
be simplified as
\begin{equation}
\begin{split}
& {P_{out}}_{\textbf{BF-IC}}(M,R,P,\rho) = \int_0^{\infty} f_{\Gamma}(\gamma) \mbox{Pr}(\mbox{outage}/\gamma)d\gamma \\
& = \frac{1}{(1+\mu)^{M-1}}\sum_{i=0}^{M-1} \binom{M-1}{i}\mu ^{i}\Gamma_{(i+1)}\left(\frac{e^R - 1}P\right).
\end{split}
\label{result_bf}
\end{equation}
The derivation of equation (\ref{result_bf}) is shown in the
appendix \ref{app1}. Note that $(1+\mu)^{M-1} = \sum_{i=0}^{M-1}
\binom{M-1}{i}\mu ^{i}$. Therefore, the result (\ref{result_bf}) can be
interpreted as the weighted average of $\Gamma_{K}\left(\frac{e^R -
1}P\right)$, which is the outage probability of a $K \times 1$ MISO
system with perfect CSIT, where $K$ varies from 1 to $M$. Therefore,
at high SNR, we expect the outage probability with BF-IC to be
dominated by the first term ($K=1$), which decays as
$\frac{1}{\textrm{SNR}}$.

The asymptotic diversity gain at infinite SNR, defined as
\begin{equation}
d = - \lim_{\textrm{SNR} \rightarrow \infty} \dfrac{\textrm{log} P_{out}}{\textrm{log} SNR},
\end{equation}
can be quantified.  From (\ref{result_bf}), using the approximation
$\Gamma{}_M(x) \simeq \frac{x^M}{M!}$ for very small $x$, we can show
that the asymptotic diversity gain of the BF-IC scheme is 1 for
imperfect CSIT, i.e.,
\begin{equation}
\textrm{Diversity Gain } d = \left\{ 
\begin{array}{ll}
1 & \textrm{for } 0 \leq \rho < 1\\
M & \textrm{for } \rho = 1
\end{array} \right.. 
\label{bfic_diversity}
\end{equation}
This result can be explained intuitively as follows. At very high SNR,
the outage probability is dominated by the error in the CSIT rather
than channel being in deep fade. However, for USPA, the asymptotic
diversity gain is $M$ independent of $\rho$. Therefore, USPA is always
better than BF-IC at high SNR. The cross-over SNR
$SNR_{cross}$($\rho$, $R$, $M$) can be obtained by equating the outage
probabilities of the two schemes: (\ref{pout_uspa}) and
(\ref{result_bf}).  Although there is no closed form expression for
cross-over SNR, it can be computed numerically. By comparing the
operating SNR with the cross-over SNR, one can switch between BF-IC
and USPA.

\subsection{Optimal Spatial Power Allocation (OSPA)}
\label{ospa-sec}
We have seen that neither beamforming nor uniform spatial power
allocation is the optimal strategy for any given $\rho$ ($0 < \rho <
1$).  We find the optimal spatial power allocation strategy that
minimizes the outage probability. Our results show that OSPA allocates
a fraction $\lambda$ of the power along the spatial mode corresponding
to the imperfect CSIT with the remaining power being equally
distributed among the other orthogonal spatial modes.

The overall outage probability is minimized by minimizing $P_{out}(\textbf{h}_{old})$,
outage probability given $\textbf{h}_{old}$, for each realization of 
$\textbf{h}_{old}$. The outage probability for a given  $\textbf{h}_{old}$ is given by 
\begin{equation}
P_{out}(\textbf{h}_{old})  =  \textrm{Pr}\left( \textbf{h}^H\textbf{Q}\textbf{h}  < \frac{e^R - 1}P\right ).
\label{ospc-pout}
\end{equation}
Using (\ref{h-hold}), $P_{out}(\textbf{h}_{old})$ can be simplified as
\begin{equation}
\textrm{Pr}\left( (\sqrt{\mu{}} \textbf{h}_{old}
+\textbf{w})\textbf{Q}((\sqrt{\mu{}} \textbf{h}_{old} + \textbf{w})^H
< \beta{} \right),
\label{ospc_form}
\end{equation}
\noindent where $\beta=\frac{e^R - 1}P(\mu+1)$ and $\mu =
\frac{\rho^2}{1-\rho^2}$.  The outage probability given by
(\ref{ospc_form}) is equivalent to the outage probability of a MISO
channel with a mean feedback of $\sqrt{\mu{}} \textbf{h}_{old}$, which
is minimized without any loss of generality by minimizing over the
fraction of the power spent in the direction of the mean feedback
\cite{costas, arie}. Rest of the power is spent equally in the M-1
orthogonal beams.

Since $\textbf{Q}$ is positive semi-definite, we have the eigenvalue
decomposition (EVD) $\textbf{Q} =
\textbf{V}\widetilde{\textbf{Q}}\textbf{V}^H$, where $\widetilde{\textbf{Q}} =
\textrm{diag}\{\lambda_1,\lambda_2,\ldots,\lambda_M\}$ is a diagonal
matrix with $\lambda_i \geq 0$ representing the power allocated to the
direction indicated by the corresponding column vector of the unitary
matrix $\textbf{V}$. It has been shown in \cite{costas} that the
unitary matrix $\textbf{V}$ that minimizes the outage probability
(\ref{ospc_form}) is of the form $\textbf{V} =
[\frac{\textbf{h}_{old}}{\sqrt{\gamma{}}},\textbf{v}_2,\textbf{v}_3,\ldots,\textbf{v}_M]$,
where $\{\textbf{v}_i\}, 2 \leq i \leq M$ is an arbitrary set of
$(M-1)$ orthonormal vectors that are orthogonal to
$\textbf{h}_{old}$. Hence, we have $\textbf{d} =
\textbf{V}^H\textbf{h}_{old} = [\sqrt{\gamma{}},0,0,\ldots,0]^T$ and $\textbf{g} =
\textbf{V}^H\textbf{w} \sim \cal{CN} (\bf{0},\bf{I})$. Thus
(\ref{ospc_form}) is simplified as
\begin{equation}
P_{out}(\textbf{h}_{old})  =  \textrm{Pr}\left( (\bf{g} +
\sqrt{\mu{}}\bf{d})^H \widetilde{\textbf{Q}} (\bf{g} +
\sqrt{\mu{}}\bf{d}) < \beta{} \right ).
\label{ospc-pout2}
\end{equation}
\begin{equation}
\begin{split}
\textrm{Let } \frac{\xi}{2} & = (\bf{g} + \sqrt{\mu{}}\bf{d})^H
\widetilde{\textbf{Q}} (\bf{g} + \sqrt{\mu{}}\bf{d}) \\
& =\bf{g}^H\widetilde{\textbf{Q}}\bf{g} + \bf{g}^H\widetilde{\textbf{Q}}\sqrt{\mu{}}\bf{d} +
   \sqrt{\mu{}}\bf{d}^H \widetilde{\textbf{Q}}\bf{g} +
   \sqrt{\mu{}}\bf{d}^H\widetilde{\textbf{Q}}\sqrt{\mu{}}\bf{d}\\
& =\sum_{i=1}^{M}\lambda_{i}\mid{}g_i\mid{} ^2
   +2\lambda_1\sqrt{\mu{}\gamma{}}\textrm{Re}(g_1)) + \lambda_1\mu\gamma\\
& =\lambda_1\lbrace[ Re(g_1) + \sqrt{\mu\gamma}]^2 + [ Im(g_1) ]^2\rbrace + \sum_{i=2}^{M}\lambda_{i}\mid{}g_i\mid{} ^2.
\end{split}
\end{equation}
Observe that $\xi$ is symmetric over $\lambda_{i},~i = 2$ to M. Hence,
there is no reason to prefer any one $\lambda_{i}$ over
others. Therefore, $\lambda_{i}$'s should be equal for $i = 2$ to
M. This observation allows the random variable $\xi$ to be expressed
in terms of $\lambda_1$ alone, using which the outage probability is
determined easily in terms of the CDF of a single random
variable. This is not explicitly used in the expressions for outage
probability in \cite{costas} (see equation (10) in
\cite{costas}). Further simplification of the outage expression based
on this observation is presented below. Denote $\lambda_{1}$ by
$\lambda{}$ for convenience.
\begin{equation}
\textrm{Tr}(\widetilde{\textbf{Q}}) = 1 \Rightarrow \lambda_{i}=
\frac{1-\lambda{}}{M-1}, \textrm{ for i}=2 \textrm{ to }M
\end{equation}
\begin{equation}
\Rightarrow \xi = \lambda A +  \frac{1-\lambda{}}{M-1} B,
\label{xi-AB}
\end{equation}
where A $= \lbrace[ \sqrt{2}Re(g_1) + \sqrt{2\mu\gamma} ]^2 + [
\sqrt{2}Im(g_1) ]^2\rbrace$ is Non-Central Chi-Square distributed with
2 degrees of freedom and non-centrality parameter $\delta =
2\mu\gamma$ and B $ = 2\sum_{i=2}^{M}\mid{}g_i\mid{}^2$ is Central
Chi-Square distributed with 2(M-1) degrees of freedom.
Observe that $\xi$ depends only on $\gamma =
\textbf{h}_{old}^H\textbf{h}_{old}$. Therefore, we denote the outage
probability for a given $\gamma{}$ and $\lambda$ as
$P_{out}(\gamma,\lambda)$, given by
\begin{equation}
P_{out}(\gamma,\lambda) = Pr(\xi < 2\beta) = F_{\xi}(2\beta),
\label{Pout-xi}
\end{equation}
where $F_{\xi}(.)$ represents the CDF of $\xi$. To complete the solution, it remains only to find the optimal value of
$\lambda{}$ for each $\gamma{}$. Consider the two extreme cases: $\rho
= 0$ and $\rho = 1$.

For $\rho = 0$, $\mu = 0$ and $\xi =
2\sum_{i=1}^{M}\lambda_{i}\mid{}g_i\mid{} ^2$ is symmetric over
$\lambda_{i}, i = 1$ to M, i.e., All the directions are identical, and
hence, equal power is spent in each direction. Therefore,
$\lambda_{opt}(\gamma) = \frac{1}{M}$, for $\rho = 0$.  As $\rho $
tends to 1, $\mu $ tends to $\infty$. Therefore, the co-efficient of
$\lambda_1$ becomes large compared to the coefficients of the other
$\lambda_i$'s, and hence, it is optimal to spend all the power in that
direction. Therefore, $\lambda_{opt}(\gamma) = 1$ for $\rho = 1$.

Consider the case of $0 < \rho < 1$. When $\gamma = 0$, $\delta =
2\mu\gamma = 0$. In this case, we get
\begin{equation}
\lambda_{opt}(\gamma = 0) = \frac{1}{M} \; \; \textrm{for any $\rho$}.
\end{equation}

As $\gamma \rightarrow \infty$, $\delta = 2\mu\gamma \rightarrow
\infty$. In this case, we get $ \lambda_{opt}(\gamma \rightarrow
\infty) = 1 \; \; \textrm{for any $\rho$}$.  Therefore, for $ 0 < \rho
< 1$, we expect $\lambda_{opt}(\gamma)$ to start from $ \frac{1}{M} $
at $\gamma = 0$ and approach 1 as $\gamma$ increases.  

The minimum outage probability for a given $\gamma{}$ is given by
$P_{out}(\gamma, \lambda_{opt}(\gamma)) = \min_{\lambda}
P_{out}(\gamma, \lambda)$, where $\lambda_{opt}(\gamma)$ is the
solution of
\begin{equation}
\frac{\partial P_{out}(\gamma, \lambda) }{\partial \lambda} = 0
\textrm{ in the range from $\frac{1}{M}$ to 1.}
\label{lam_opt}
\end{equation}

Expressing $P_{out}(\gamma, \lambda)$ as
\begin{equation}
\begin{split}
P_{out}(\gamma, \lambda) = \mbox{Pr} \left( \lambda A +  \frac{1-\lambda{}}{M-1} B < 2\beta \right)&\\
= \int_0^{\frac{2\beta}{\lambda}} f_A(a) F_B\left(\dfrac{(2\beta - \lambda{}a)(M-1)}{1-\lambda{}}\right) da,&
\end{split}
\end{equation}
equation (\ref{lam_opt}) can be simplified as
\begin{equation}
\begin{split}
\int_0^{\frac{2\beta}{\lambda}} f_A(a) \textrm{exp}\left(
\frac{(M-1)\lambda{}a}{2(1-\lambda{})}\right)(2\beta -
\lambda{}a)^{(M-2)}&(2\beta - a) da = 0.
\end{split}
\label{lam_constraint}
\end{equation}
Although a closed form expression for $\lambda_{opt}(\gamma{})$ does
not appear to be available, it can be determined numerically by a
one-dimensional numerical search over the range.  The overall outage
probability can then be determined by averaging
$P_{out}(\gamma,\lambda_{opt}(\gamma))$ over $\gamma$.

\vspace{-6mm}
\section{Long-Term Power Constraint}
\label{ltpc}
Achieving minimum outage probability under a long-term power
constraint involves power allocation in both spatial and temporal
domains. For a given feedback SNR $\gamma{}$, and a corresponding fixed power
allocation policy, the problem of minimizing the outage probability
can be formulated as
\begin{equation}
\min_{\bf{Q}}\textrm{Pr}\left( \textbf{h}^H\textbf{Q}\textbf{h}  < \frac{e^R -
  1}{Pp(\gamma{})}\right ).
\end{equation}
This is equivalent to minimizing $P_{out}(\gamma,p(\gamma),\lambda)$,
given by
\begin{equation}
\begin{split}
P_{out}(\gamma, p(\gamma{}), \lambda) = \mbox{Pr} \left( \lambda A +
\frac{1-\lambda{}}{M-1} B < \frac{2\beta}{p(\gamma{})} \right) = F_{\xi}\left(\frac{2\beta}{p(\gamma)}\right)
\end{split}
\label{Pout-xi-pc}
\end{equation}
over $\lambda$, fraction of the power spent in the direction of the
imperfect feedback. $\lambda_{opt}(\gamma{}, p(\gamma{}))$ is the
solution of $\displaystyle{\frac{\partial P_{out}(\gamma, p(\gamma{}), \lambda) }{\partial
\lambda} = 0}$ and will range from $\frac{1}{M}$ to 1.
The optimal temporal power control policy $p(\gamma)$ minimizes
\[
E_{\gamma}[P_{out}(\gamma,p(\gamma),\lambda_{opt}(\gamma{},
p(\gamma{})))] = \int_0^{\infty} f_{\Gamma}(\gamma) P_{out}(\gamma, p(\gamma{}), \lambda_{opt}(\gamma, p(\gamma{})))d\gamma,
\]
subject to the power constraint: 
\begin{equation}
\int_0^{\infty} f_{\Gamma}(\gamma) p(\gamma{})d\gamma = 1.
\label{power-cons}
\end{equation}

However, finding optimal $p(\gamma{})$ and the corresponding
$\lambda_{opt}(\gamma, p(\gamma{})) $ is difficult, since we do not
have closed form expression for $\lambda_{opt}(\gamma, p(\gamma{}))
$. Therefore, based on the intuition from the results for the
short-term power constraint, the suboptimal schemes BF-IC with
temporal power control and USPA with temporal power control are
considered and analyzed. 

In USPA, the power is distributed equally among the orthogonal
independent transmit directions, i.e, $\lambda = \frac{1}{M}$ or
$\bf{Q} = \frac{\bf{I}}{M}$. Therefore, the outage probability for a
given $\gamma{}$ and the corresponding $p(\gamma{})$ in equation
(\ref{Pout-xi-pc}) is simplified as
\begin{equation}
\begin{split}
P_{out}(\gamma, p(\gamma{})) = \mbox{Pr} \left( A + B <
\frac{2M\beta}{p(\gamma{})} \right) =
F_{(nc-\chi{}^2,2M,\delta{})}\left( \frac{2M\beta}{p(\gamma{})}
\right).
\end{split}
\label{Pout-uspa-pc-gamma}
\end{equation}

From calculus of variations \cite{brunt} (using Theorem 4.2.1 in
\cite{brunt}), the temporal power control function that minimizes the
outage probability with USPA can be shown to satisfy:
\begin{equation}
k_1 = \left(\frac{2M\beta}{p^2(\gamma{})}\right)f_{(\text{nc-}\chi^2,2M,\delta)}\left(\frac{2M\beta}{p(\gamma{})}\right),
\label{p_gamma_cons}
\end{equation}
where $k_1$ is a constant chosen such that $p(\gamma)$ satisfies the
power constraint (\ref{power-cons}) and is non-negative. Finally,
$p(\gamma)$ is determined numerically from equations
(\ref{power-cons}) and (\ref{p_gamma_cons}).

In BF-IC, the spatial power allocation scheme is fixed such that the
power is spent in only one direction corresponding to the imperfect
CSIT, i.e., $\lambda = 1$. Therefore, we have 
\begin{equation}
\begin{split}
P_{out}(\gamma, p(\gamma{})) = \mbox{Pr} \left(A < \frac{2\beta}{p(\gamma{})}
\right) =
F_{(\text{nc-}\chi^2,2,\delta)}\left(\frac{2\beta}{p(\gamma{})}\right).
\end{split}
\label{Pout-xi-bfpc}
\end{equation}
Again, using calculus of variations, the termporal power control
function that minimizes the outage probability with BF-IC can be shown
to satisfy
\begin{equation}
k_2 = \left(\frac{2\beta}{p^2(\gamma{})}\right)f_{(\text{nc-}\chi^2,2,\delta)}\left(\frac{2\beta}{p(\gamma{})}\right),
\label{p_gamma_cons2}
\end{equation}
where $k_2$ is a constant, chosen such that $p(\gamma)$ satisfies the
power constraint (\ref{power-cons}) and is non-negative. $p(\gamma)$
can be obtained numerically as before using equations (\ref{power-cons})
and (\ref{p_gamma_cons2}).

\vspace*{-6mm}
\section{Effect of Imperfect CSIR}
\label{imcsir}
\vspace*{-2mm} 
We assume the training and MMSE channel estimation
model as in \cite{sk-ba}. $M$ training symbols are transmitted at the
start of each $T$ symbol block with the $i^{th}$ training symbol being
transmitted only from the $i^{th}$ antenna. The MMSE estimate of CSI
($\hat{\textbf{h}}$) is:
 \begin{equation}
{\mathbf{\hat{h}}} = \frac{\sqrt{P_t / M}}{P_t/M + 1} \left( \sqrt{\frac{P_t}{M}} {\mathbf h} + {\mathbf n} \right),
\label{mmse-est}
\end{equation}
where $P_t$ is the total power used for training, and ${\bf n}$ is the
additive white Gaussian noise vector corresponding to the $M$ training
symbols. Let $P_d$ be the power used during data transmission period
per symbol. $P_t$ and $P_d$ are related by the equation: $P_t +
P_d(T-M) = PT$. Let $\sigma_E^2$ denote the estimation error variance,
i.e., $Cov(\textbf{e}) = \sigma_E^2\textbf{I}_{M \times M},$ where
$\textbf{e} = \textbf{h} - \hat{\textbf{h}}$. It can be shown that
$\sigma_E^2 = \frac{M}{P_t+M}.$ The CSIR is $\hat{\textbf{h}}$. The
CSIT, which is a delayed version of the CSIR is
$\hat{\textbf{h}}_{old}$, which is the MMSE estimate of
$\textbf{h}_{old}$. Using (\ref{h-hold}) and (\ref{mmse-est}), the
correlation coefficient $\rho_e$ between the CSIT
($\hat{\textbf{h}}_{old}$) and CSIR ($\hat{\textbf{h}}$) can be
obtained as $ \rho_e = \frac{P_t}{P_t+M}\rho$.  Observe that $\rho_e$
can at most be $\rho$ (for very large training power) and is less than
$\rho$ for moderate values of training power. 

Given the MMSE estimate of the CSI ($\hat{\textbf{h}}$) at the
receiver, the mutual information of the BF-IC scheme after accounting
for the training period can be lower bounded using the result in
\cite{LapSha02}. A similar mutual information lower bound can be
obtained for the USPA scheme using the results in
\cite{yoo-ag,sk-ba}. This lower bound on the mutual information is
given by:
\begin{eqnarray} \textrm{I}(\textbf{x};y|\hat{\textbf{h}},\hat{\textbf{h}}_{old}) \ge \frac{T-M}{T}\textrm{log}\left(1 + \frac{P_d}{1+\sigma{}_E^2P_d}\hat{\textbf{h}}^H\textbf{Q}\hat{\textbf{h}}\right).
  \label{I-imperfectCSIR}
\end{eqnarray}

Defining $\hat{\textbf{h}}_{sc} = \frac{1}{\sqrt{(1-\sigma_E^2)}}
\hat{\textbf{h}}$ such that $Cov(\hat{\textbf{h}}_{sc}) =
\textbf{I}_{M \times M}$, the lower bound on the mutual information
(\ref{I-imperfectCSIR}) can be written as
\begin{equation}
 \begin{split} \textrm{I}(\textbf{x};y|\hat{\textbf{h}},\hat{\textbf{h}}_{old}) \ge& \frac{T-M}{T}\textrm{log}\left(1 + P' \hat{\textbf{h}}_{sc}^H\textbf{Q}\hat{\textbf{h}}_{sc}\right),\\
 \textrm{where } P' =& P_d\frac{1-\sigma_E^2}{1+\sigma{}_E^2P_d}.
 \end{split}
 \label{Isc-imperfectCSIR}
\end{equation}
Substituting the value of $\sigma_E^2$ obtained for the training model, we get $P' = \frac{P_dP_t}{P_t+MP_d+M}$.

For USPA, $\textbf{Q} = \frac{\textbf{I}_{M \times
M}}{M}$. Clearly, the lower bound on mutual information above
becomes equivalent to a system with perfect CSIR, but with different
values of average SNR ($P'$) and rate ($R'$). Therefore, the outage
probability is upper bounded as follows:
\begin{equation}
\begin{split}
	{P_{out}}_{\textbf{USPA}}(M,R,P') \le& \mbox{Pr}\left(\textrm{log}\left(1 + P'\frac{ \hat{\textbf{h}}_{sc}^H\hat{\textbf{h}}_{sc}}{M}\right) < R\frac{T}{T-M}\right)
	= \Gamma{}_M\left( \frac{e^{R'} - 1}{P'/M}\right),
	\end{split}
\end{equation}
\begin{equation}
	\textrm{where } P' = \frac{P_dP_t}{P_t+MP_d+M}, R' = R\frac{T}{T-M}
\label{p'r'}
\end{equation}
The asymptotic diversity gain of USPA with imperfect CSIR remains
$M$. Furthermore, the SNR gap between the perfect and imperfect CSIR
cases can be significantly reduced by choosing value of $P_t$ or $P_d$
that maximizes $P'$ under the constraint $P_t + P_d(T-M) = PT$
\cite{sk-ba}.  

In BF-IC, the transmit covariance matrix is
%\begin{equation}
$\displaystyle{\textbf{Q} = \frac{\hat{\textbf{h}}_{old}\hat{\textbf{h}}_{old}^H}{\hat{\textbf{h}}_{old}^H\hat{\textbf{h}}_{old}} =
\frac{\hat{\textbf{h}}_{old,sc}\hat{\textbf{h}}_{old,sc}^H}{\hat{\textbf{h}}_{old,sc}^H\hat{\textbf{h}}_{old,sc}}},$
% \end{equation}
where $\hat{\textbf{h}}_{old,sc}$ is a scaled version of
$\hat{\textbf{h}}_{old}$ with identity covariance matrix.  Again, the
lower bound on the system with imperfect CSIR is equivalent to the
system with perfect CSIR with the parameters: average SNR ($P'$) and
rate ($R'$) given by (\ref{p'r'}) and $\rho_e$. Following the
simplifications as in Section \ref{bficperfect} and the appendix, we
get
\begin{equation}
{P_{out}}_{\textbf{BF-IC}}(M,R',P',\rho) \le \frac{1}{(1+\mu')^{M-1}}\sum_{i=0}^{M-1} \binom{M-1}{i}(\mu') ^{i}\Gamma_{(i+1)}\left(\frac{e^{R'} - 1}{P'}\right),
\end{equation}
where $\mu' = \frac{\rho{}_e^2}{1-\rho{}_e^2}$, and $P'$, $R'$ are
given by equation (\ref{p'r'}). 

\section{Results \& Observations}
\label{results}
\begin{figure}[ht]
	\begin{center}
		\includegraphics[width=4.8in]{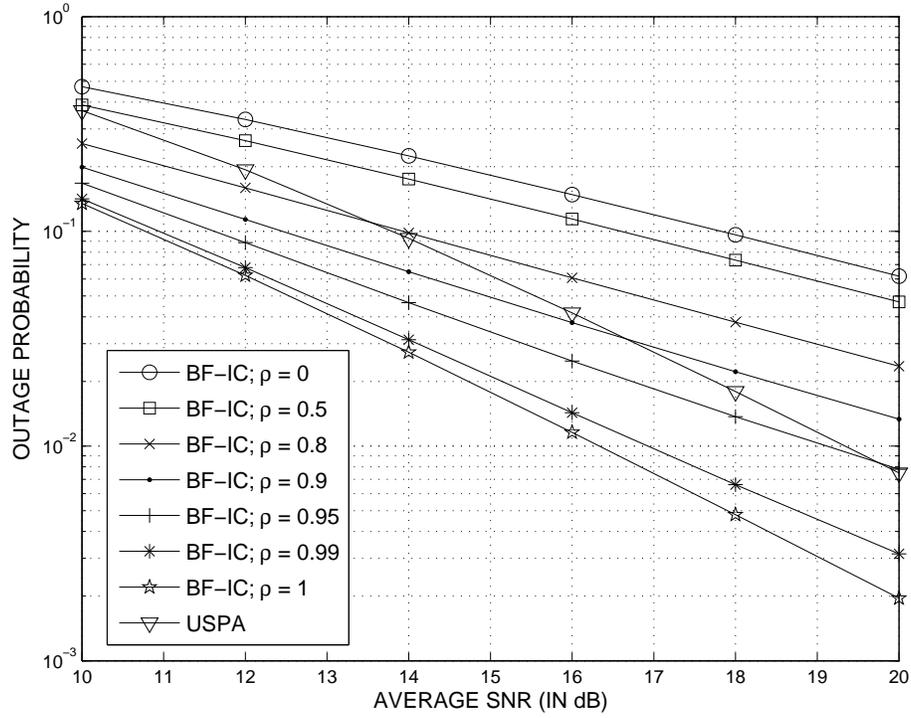}
	\end{center}
	\caption{Outage probabilities for Beamforming using imperfect
	CSIT (BF-IC) for various values of $\rho$, and uniform spatial
	power allocation (USPA) for $M = 2$ and $R = 2$
	nats/s/Hz. Cross-over SNR is the SNR at which
	USPA and BF-IC have the same outage probability.}
  \label{bf_0_1}
\end{figure}

\begin{figure}[ht]
	\begin{center}
		\includegraphics[width=4.8in]{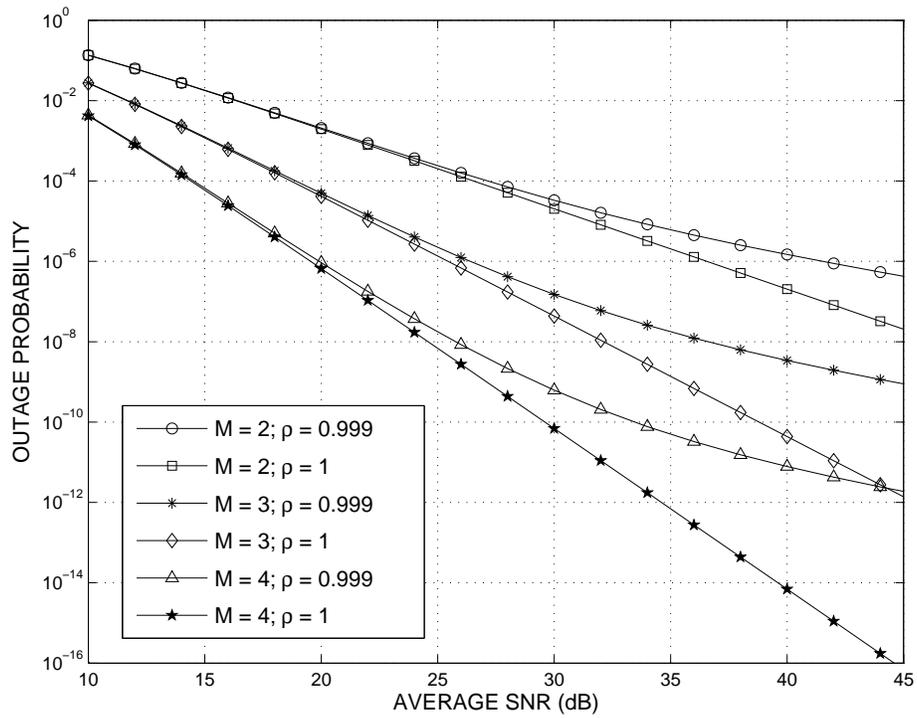}
	\end{center}
	\caption{Outage probability with BF-IC for various values of $M$ for $\rho{} = 0.999$ and beamforming for $\rho = 1$ and $R = 2$ nats/s/Hz}
	\label{M_2_4_rho_0999}
\end{figure}

The rate of transmission (R) is chosen to be 2 nats/s/Hz throughout
this section. Fig. \ref{bf_0_1} shows the performance of USPA
(\ref{pout_uspa}) and BF-IC (\ref{result_bf}) for different values of
feedback delay captured by $\rho$. BF-IC is better at lower SNRs and
worse at high SNRs when compared to USPA for any $\rho < 1$.
Fig. \ref{M_2_4_rho_0999} shows the diversity gain of USPA and BF-IC
for different number of transmit antennas (M) for $\rho = 0.999$. USPA
does not require feedback and has a diversity gain of $M$, where as at
high SNR, the outage probability with BF-IC is dominated by the error
in CSIT. Thus, the diversity gain of BF-IC scheme is equal to 1, for
any non zero delay in the feedback and any number of transmit
antennas. Hence, USPA outperforms BF-IC at high SNR for all values of
$\rho < 1$. Cross-over SNR is defined as the SNR after which USPA
outperforms BF-IC. It can be seen from Fig. \ref{bf_0_1} that the
cross-over SNR is a monotonically increasing function of $\rho$.

\begin{figure}[ht]
	\begin{center}
		\includegraphics[width=4.8in]{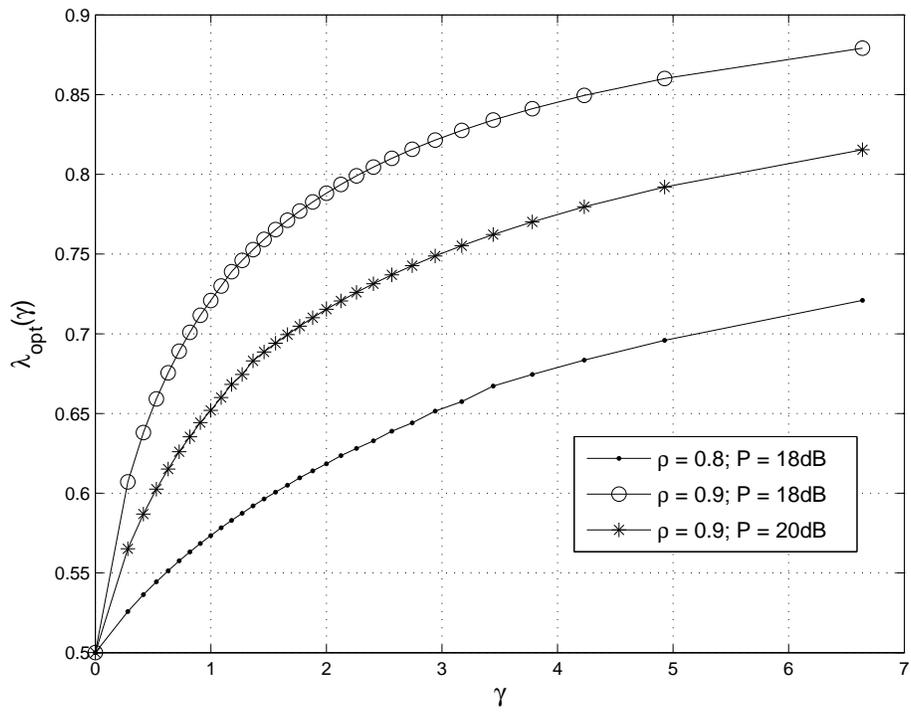}
	\end{center}
	\caption{Fraction of the power in the direction of
	imperfect CSIT $\lambda_{opt}(\gamma)$ for different values of
	$\rho$ and $P$; $M=2$ and $R=2$
	nats/s/Hz.}
	\label{lam_optfig}
\end{figure}

\begin{figure}[ht]
	\begin{center}
		\includegraphics[width=4.8in]{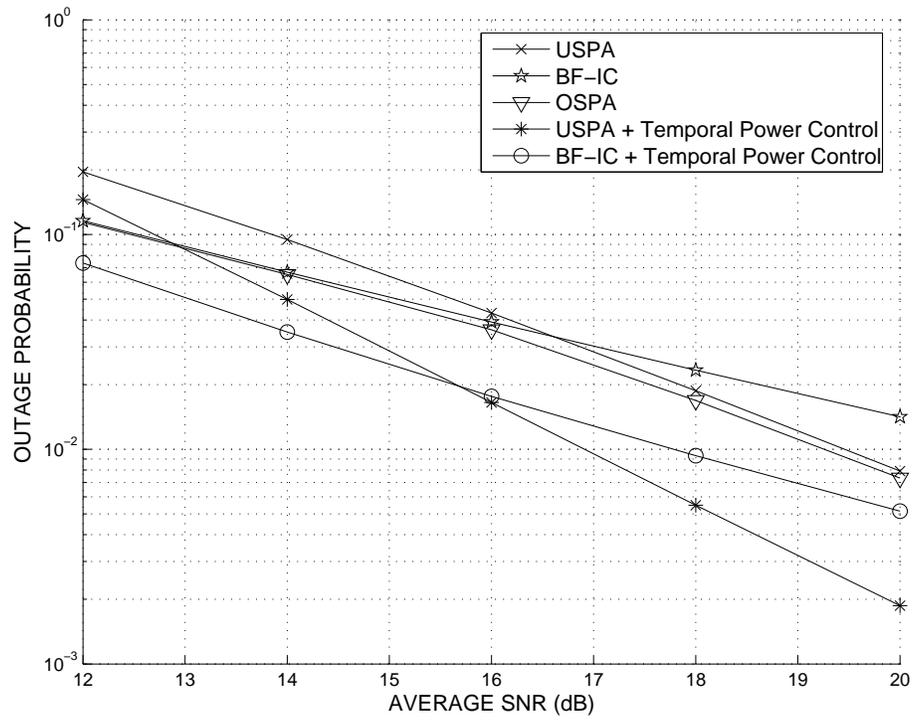}
	\end{center}
	\caption{Outage Probabilities for uniform spatial power
	allocation (USPA) and beamforming using imperfect CSIT (BF-IC)
	with and without temporal power control, and optimal spatial
	power allocation (OSPA) for $\rho = 0.9$; $M=2$ and $R=2$ nats/s/Hz.}
	\label{compare_spc_09}
\end{figure}

Fig. \ref{lam_optfig} shows $\lambda_{opt}(\gamma)$, the fraction of
power spent in the direction of imperfect CSIT, as a function $\gamma$
for the OSPA scheme. Observe that $\lambda_{opt}(\gamma)$ is larger
for higher values of $\rho$ implying that when the quality of feedback
is higher, more power is spent in the direction of
feedback. Fig. \ref{compare_spc_09} compares the outage probability of
BF-IC and USPA with OSPA for $\rho = 0.9$. We observe that (a) OSPA
provides negligible gain in performance, (b) OSPA is computationally
complex as it requires the transmitter to compute the optimal value of
$\lambda$ for each value of feedback SNR and adapt the power in the
spatial modes correspondingly, and (c) OSPA requires an estimate of
$\rho$ to determine $\lambda_{opt}(\gamma)$ and any mismatch between
the estimated value and the actual value will hurt the performance. On
the other hand, USPA and BF-IC do not require any estimate of $\rho$
and are very simple. Therefore, we suggest switching between BF-IC and
USPA by comparing the operating average SNR with the cross-over SNR.
For a given average SNR, it is also possible to choose between USPA
and BF-IC based on the instantaneous feedback SNR $\gamma$ (instead of
switching based on the average SNR irrespective of
$\gamma$). Equations (\ref{pout_uspa}) and (\ref{pout-bf-gamma}) are
the outage probabilties of USPA and BF-IC for a given
$\gamma$. However, we know that switching based on average SNR is
already very close to the performance of OSPA. Therefore, the possible
improvement due to switching based on instantaneous SNR (instead of
average SNR) is very small.

Fig. \ref{compare_spc_09} also shows the performance of BF-IC and USPA
with the corresponding optimal temporal power control. As in the case
of the short-term power constraint, temporal power control with BF-IC
is better for low SNR and temporal power control with USPA is better
at high SNR. The cross-over SNR is slightly lower with temporal power
control.

\begin{figure}[ht]
	\begin{center}
		\includegraphics[width=4.8in]{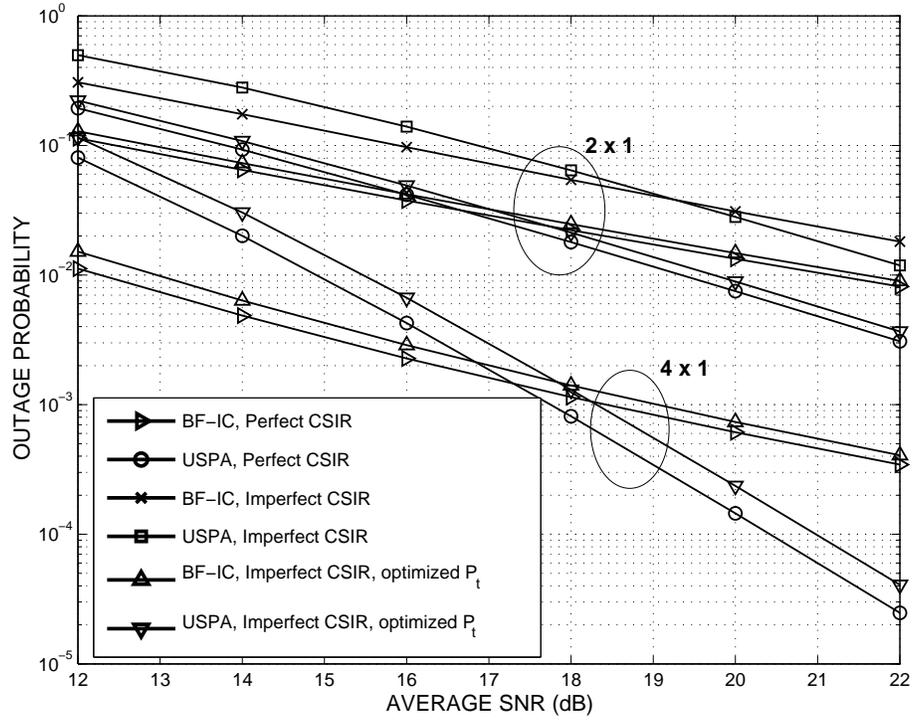}
	\end{center}
	\caption{Outage probabilities for Beamforming using imperfect
	CSIT (BF-IC) for $\rho = 0.9$, and uniform spatial power
	allocation (USPA) for $M = 2, 4$ and $R = 2$ nats/s/Hz. Outage
	for $M = 4$ with imperfect CSIR is plotted only for the
	optimized trianing power case.}
  \label{pout_bf_uspa}
\end{figure}

Fig. \ref{pout_bf_uspa} compares the outage probability of BF-IC and
USPA for perfect CSIR with BF-IC and USPA for imperfect CSIR for $\rho
= 0.9$. Both $M = 2$ and $M=4$ are considered. $T$ is chosen to be
100. Outage probability for two cases: (a) Preamble power same as data
power ($P_d = P_t$) and (b) optimized preamble power is
considered. Note that optimal power chosen for USPA is used as it is
for BF-IC. The results suggest that the loss in performance due to
imperfect CSIR is not significant for both USPA and BF-IC, if the
power is chosen appropriately.

\vspace*{-5mm}
\section{Summary}
\label{conc}
The problem of minimum outage transmission for a MISO system with M
transmit antennas with delayed feedback is considered. The delay in
the feedback is captured by $\rho$, the correlation coefficient
between delayed CSIT and perfect CSIR. For a short-term power
constraint, we derive an analytic expression for the outage
probability of beamforming using imperfect CSIT, where the power is
spent in only one direction corresponding to the imperfect CSI
available with the transmitter and compare it with that of USPA, where
the power is distributed equally among the M orthogonal and
independent transmit directions.  We also determine the optimal
transmit strategy, i.e., OSPA, that minimizes the outage probability
numerically. OSPA involves allocating a fraction of the power in the
direction of the imperfect CSIT and the rest of the power is equally
distributed among the $M-1$ orthogonal and independent transmit
directions. Results show that, for any $\rho < 1$, BF-IC is better at
low SNR and worse at high SNR when compared to USPA. Furthermore, the
asymptotic diversity gain for BF-IC is equal to 1 for any $\rho < 1$,
independent of the number of transmit antennas. BF-IC is close to
optimal at low SNR, while USPA is close to optimal at high SNR, i.e.,
OSPA does not improve the outage probability significantly compared to
switching between BF-IC and USPA depending on the average SNR. The
cross-over SNR can be determined numerically by equating the outage
probabilities of BF-IC and USPA schemes. For the long-term power
constraint, where the transmit power is varied with time based on the
feedback SNR, we numerically evaluate the outage probabilities and
show again that BF-IC is better at low SNR, while USPA is better at
high SNR. Finally, we show that the performance loss due to imperfect
CSIR is minimal if the training power is chosen appropriately.

\vspace*{-6mm}
\appendix[Derivation of equation (\ref{result_bf})]
\label{app1}
\vspace*{-6mm}
\begin{equation}
F_{(\text{nc-}\chi^2,2M,\delta)}(y) = \sum_{k=0}^{\infty}
\frac{\left(\frac{\delta}{2}\right)^k e^{-\frac{\delta}{2}}}{k!}
F_{\chi^2,2M+2k}(y),
\label{ncx2cdf}
\end{equation}
where $F_{\chi^2,2M+2k}(.)$ is the cdf of a central $\chi^2$ random
variable with $2M+2k$ degrees of freedom.  Using (\ref{ncx2cdf}),
(\ref{pout-bf-gamma}) and substituting $\delta = 2\mu{}\gamma{}$,
${P_{out}}_{\textbf{BF-IC}}(M,R,P,\rho)$ is simplified as
\begin{equation}
\begin{split}
{P_{out}}_{\textbf{BF-IC}}(M,R,P,\rho) & = \int_0^{\infty}
f_{\Gamma}(\gamma) \sum_{k=0}^{\infty} \frac{(\mu{}\gamma{})^k
e^{-\mu{}\gamma{}}}{k!} F_{\chi^2,2+2k}(2\beta{})d\gamma\\ & =
\sum_{k=0}^{\infty} \frac{\mu{}^k}{k!}
F_{\chi^2,2+2k}(2\beta{})\int_0^{\infty} f_{\Gamma}(\gamma)\gamma{}^k
e^{-\mu{}\gamma{}}d\gamma \\ & = \sum_{k=0}^{\infty}
\frac{\mu{}^k}{k!}\int_0^{\beta{}}\frac{x^ke^{-x}}{k!}dx
\frac{(M+k-1)!}{(M-1)!(1+\mu{})^{(M+k)}}\\ & =
\frac{1}{(1+\mu)^{M}}\int_0^{\beta}e^{-x}g(x)dx,
\end{split}
\label{pout_gofx}
\end{equation}
\begin{equation}
\textrm{where } g(x) = \sum_{k=0}^{\infty} \frac{\binom{M+k-1}{k}}{k!}\left(\frac{\mu x}{1+\mu}\right)^{k}.
\end{equation}
LEMMA. For any $m,n>0$,
\begin{equation}
\binom{m+n}{m} = \binom{m+n}{n} = \sum_{i=0}^{min(m,n)}\binom{m}{i}\binom{n}{i}.
\label{lemma-1}
\end{equation}
Proof: Using symmetry in $m$ and $n$, we assume $m<n$ without any loss
of generality. Observe that the L.H.S is the number of ways to chose
$m$ objects out of $m+n$. This can also be calculated by separating
the $m+n$ objects in to 2 sets with sizes $m$ and $n$ and choosing $i$
objects from the first set and choosing $n-i$ objects from the second
set and varying $i$ from 0 to $m$. Therefore,
\[
\binom{m+n}{m} =  \sum_{i=0}^m\binom{m}{i}\binom{n}{n-i} = \sum_{i=0}^{min(m,n)}\binom{m}{i}\binom{n}{i}.\hfill \boxempty 
\]
Using the above lemma, $g(x)$ can be simplified as 
\[
g(x)  = \sum_{k=0}^{\infty}
\sum_{i=0}^{min(k,M-1)}\binom{M-1}{i}\binom{k}{i}\frac{1}{k!}\left(\frac{\mu
x}{1+\mu}\right)^{k} 
\]

\[  = \sum_{i=0}^{M-1}
\sum_{(k-i)=0}^{\infty}\binom{M-1}{i}\frac{1}{i!(k-i)!}\left(\frac{\mu
x}{1+\mu}\right)^{k} 
\]
\[  = \sum_{i=0}^{M-1}
\frac{\binom{M-1}{i}}{i!}\left(\frac{\mu x}{1+\mu}\right)^i
\sum_{(k-i)=0}^{\infty}\frac{1}{(k-i)!}\left(\frac{\mu
x}{1+\mu}\right)^{k-i}
\]
\[  = e^{\left(\frac{\mu
x}{1+\mu}\right)}\sum_{i=0}^{M-1}
\frac{\binom{M-1}{i}}{i!}\left(\frac{\mu x}{1+\mu}\right)^{i}.
\]
After substituting for $g(x)$ in (\ref{pout_gofx}), we get
\begin{equation}
\begin{split}
{P_{out}}_{\textbf{BF-IC}}(M,R,P,\rho) & =
\frac{1}{(1+\mu)^{M}}\sum_{i=0}^{M-1}
\frac{\binom{M-1}{i}}{i!}\mu{}^i\int_0^{\beta}
e^{-\left(\frac{x}{1+\mu}\right)}\left(\frac{x}{1+\mu}\right)^{i}dx \\
& = \frac{1}{(1+\mu)^{M-1}}\sum_{i=0}^{M-1} \binom{(M-1)}{i}\mu
^{i}\Gamma_{(i+1)}\left(\frac{e^R - 1}P\right).
\end{split}
\end{equation}

\vspace*{-2mm}
\bibliographystyle{IEEE}
\bibliography{tcom07_rev2}

\newpage

\end{document}